\documentclass[aps,a4paper,showpacs,showkeys,preprint]{revtex4}
\usepackage{amsmath}
\usepackage{amssymb}
\usepackage{graphics}
\usepackage{psfrag}
\usepackage{epsfig}
\usepackage{hyperref}
\usepackage{ulem}
\usepackage{color}

\begin{document}

\title{Ising spin glass in a random network with a gaussian random field} \author{R. Erichsen Jr, https://pt.overleaf.com/project/5f7f0df369a1bf00014a6d7dA. Silveira and S. G. Magalhaes} \affiliation{Instituto de F\'{\i}sica, Universidade  Federal do Rio Grande do Sul, Caixa Postal 15051, 91501-970 Porto Alegre, RS, Brazil}

\date{\today}

\begin{abstract}

We investigate the thermodynamic phase transitions of the joint presence of Spin Glass (SG) and Random Field (RF) using a random graph model
that allows to deal with the quenched disorder.  
Therefore, the connectivity becomes a controllable parameter in our theory allowing to answer 
whether and which are the differences between this 
description and
the mean field theory
i. e., the fully connected theory.  We have considered the Random Network Random Field Ising Model (RNRFIM) where the spin exchange interaction, as well as the RF, are random variables following a gaussian distribution. The results were found within the replica symmetric (RS) approximation whose stability is 
obtained using the
Two-replica method. 
This also puts our work in the context of a broader discussion 
which is the RS stability as a function of the connectivity. 
In particular, 
our results show that for small connectivity there is a region at zero temperature where the RS solution remains stable  
above a given value of the magnetic field no matter the strength of RF. Consequently, our results 
 show important differences
with the crossover between the RF and SG regimes predicted by the fully connected theory.

\end{abstract}



\preprint{IF-UFRGS 2020}

\thispagestyle{empty}

\maketitle

\section{Introduction}

The issue of disorder in spin systems is an inexhaustible source of
problems. Two manifestations of disorder, spin glass (SG) and random
fields (RFs), illustrate how rich this research area can be
\cite{Younglivro}.  Undeniably, the corresponding theory has been
recognized for its conceptual richness, awakening interest and
providing knowledge not only in physics, but also in other fields such
as information theory and computer science, among others.  Therefore,
one can expect that the joint presence of SG and RFs can bring plenty
of fascinating possibilities.  Most importantly, it is not only a
theoretical possibility.  In fact, the joint presence of SG and RF has
been suggested in physical systems as distinct as ferro and
antiferroelectric crystals such as Rb$_{1-x}$(NH$_4$)$_x$H$_2$PO$_4$
\cite{Slak1984}, the diluted antiferromagnet Fe$_x$Zn${_{1-x}}$F$_2$
\cite{Belanger1993} and the diluted ferromagnet LiHo$_x$Y$_{1-x}$F$_4$
\cite{Wu1993,Tabei2006}. The first case being the realization of the
electrical equivalent of a SG with pseudospins degree of freedom. In
the three cases, the applied magnetic field couples with Ising spins
(or pseudospins).  For Rb$_{1-x}$(NH$_4$)$_x$H$_2$PO$_4$ and
LiHo$_x$Y$_{1-x}$F$_4$, the field transverse to the Ising direction
leads to a quantum phase transition
\cite{Pirc1987,Morais2016,Magalhaes2017}. The diversity of systems and
scenarios to describe them can anticipate that the theoretical
description of the joint presence of SG and RF might be a ground
favoring the rise of conceptual and methodological novelties.

A pertinent question is the extent to which the mean field theory can
provide a realistic description of the joint presence of SG and RF.
As an example but which may allow more general conclusions, one can
mention the mean field description of the Fe$_x$Zn$_{1-x}$F$_2$.  
Although Fe$_x$Zn$_{1-x}$F$_2$ has short-range interactions, the mean
field description \cite{Soares1994}, i.e., the infinite-range
Sherrington-Kirkpatrick (SK) model \cite{SK1975} with a gaussian
distribution for the RF  to describe some aspects of the behavior of the the mentioned system depending on the parameter $\Delta/J$ where $\Delta$
and $J$ are the RF and random spin exchange interaction variances,
respectively.  Interestingly, it was proposed that there is 
a crossover between the SG
and RF regimes by varying $\Delta/J$ or by varying $T/J$ ($T$ is
the temperature) for a fixed $\Delta/J$. This crossover is
described by $\tau \equiv T_0-T\sim h_0^{2/\phi}$ ($T_0$ is the
freezing temperature without field). The crossover has the
exponent $\phi=1$ when the RF regime dominate and $\phi=3$ as given by
the Almeida-Thouless (AT) line \cite{AT}, i.e., the line that signals the limit of stability of the replica symmetric (RS) solution.  
In addition, the mean field theory predicts that the
typical behavior of the AT line is robust for any value of $\Delta/J$
as $h_0$  increases,
i.e., there is an exponentially
small region with the SG non-trivial ergodicity breaking even for
$\Delta\gg J$. However, one can asks whether this crossover description is robust. In this direction of
investigation, a more specific question can be raised. For
instance, what happens in the limit of high magnetic fields? The mean
field description in Ref. \cite{Soares1994} predicts that the
behavior in that limit is given by the AT line. This scenario is
particularly not easy to reconcile with the direction that the debate
on the existence of the AT line for disordered spins SG systems with
short-range interaction has taken (see, for instance, \cite{Parisi1979,Fisher1985,Young2004, Young2008,Mydosh2015}).  
Nevertheless, the difficulties
in providing answers to these questions are in fact the difficulties in
describing disordered spins systems with short-range
interactions. This puts the need for alternative approaches that can
bring substantial improvements over the mean field description.

Our proposal is
to use random networks. The main reason is that these networks do
allow that the coordination number, i.e., the network  connectivity becomes a controllable parameter of the theory
\cite{Mo98,ETM,Neri,skantzos}.  Thereby, one can interpolate between
the limit of high connectivity that would be closer to the usual mean field 
theory  (called from now on, the fully connected 
theory)
until the situation with a spin with very few connections to
other spins.  Although this limit is not equivalent to treating the problem with short-range interactions, it can certainly highlight, at least, the limitations of the fully connected description.
Actually, this approach has been already used in
Ising spin system with a RF.  The Random Network Random Field Ising
Model (RNRFIM) was developed in Ref.  \cite{Doria} to study the
ferromagnetic (FM) to paramagnetic (PM) transition in networks with
non-uniform, finite averaged connectivity. There, the existing
couplings were uniformly ferromagnetic and the disorder was restricted
to the existence or not of a bond between two given sites. 
The results there showed that the existence of a
tricritical point, when the RF distribution is
discrete,  is very dependent on the connectivity. In fact, the tricritical point 
tends
to
disappear when connectivity is very small. This result is in evident
contrast with the fully connected
theory for the RF Ising model.
\cite{Aharony1978}.

In the present work, we use the RNRFIM
where, besides the presence of
a RF with a gaussian distribution, the spin couplings are also
disordered following the same kind of distribution. This type of choices
allows us to investigate not only the SG to PM phase transition but also the SG to FM
phase transition in the presence of a RF.  Since the RF couples
with the local magnetic moments, the replica symmetric (RS)
Edwards-Anderson SG order parameter will be induced whenever a RF is applied, turning out to be not useful to localize the SG
transition. Therefore, 
it is inescapable to test the
RS stability 
to locate the onset of non-trivial ergodicity associated to the SG
transition \cite{Magalhaes2011}. 
In order to accomplish that, we
use
the Two-replica method \cite{TwoReplica} which is 
quite suitable to our approach, since it
allows to obtain the limits of stability of the RS approximation
using the RS calculations themselves. 
In particular, having obtained the limits of RS stability
, i.e., the AT line, 
and counting connectivity and RF variance
as controllable parameters, we can check any crossover between SG and
RF regimes in low and high connectivity scenarios 
when a magnetic
field is applied.   
For completeness, we also investigate effects of
connectivity on the non-linear susceptibility $\chi_3$. This quantity is a well established
fingerprint of the SG transition \cite{binderyoung}.
It is known  that $\chi_3$ is strongly affected by the RF in the limit of the fully connected random network \cite{Morais2016,Magalhaes2017}.  
Therefore, it is also an interesting issue how 
the RF affects in the $\chi_3$ at low connectivity.

Lastly, we remark that there exist other approaches to deal with  finite connectivity in spins disordered problems, such as the cavity method (see, for instance, \cite{Ricci1,Ricci2}). However, we focus mainly in the RS approximation for which the random network is quite suitable.  The development of a replica symmetry breaking theory for the random network with finite connectivity for the SG problem with RF is beyond the objective of this paper.

The paper is organized as follows: in Sec. II, the free energy and
order parameter are obtained using finite connectivity within the RS
scheme. The two-replica method, employed to localize the AT line is
explained in this section. Sections III and IV present the theoretical
results and the results obtained from numerical simulations,
respectively.  Section V offers concluding remarks.

\section{The model}

The hamiltonian is an extension of RNRFIM which has two-sites 
disordered interaction
and local random field to single site interaction terms,
\begin{align}
  H=-\sum_{i,j<i}\sigma_ic_{ij}J_{ij}\sigma_j -\sum_i
  h_i\sigma_i\,,
  \label{hamiltonian}
\end{align}
where $i=1\cdots N$ and $\sigma_i=\pm 1$ are canonical Ising spin
variables. The connectivity variables $c_{ij}$ are independent,
identically distributed random variables (i.i.d.r.v.) chosen according
the probability distribution
\begin{align}
  p\left(c_{ij}\right)=\frac{c}{N}\delta_{c_{ij},1}
  +\left(1-\frac{c}{N}\right)\delta_{c_{ij},0}\,,
  \label{cdistr}
\end{align}
where $c\in\mathbb{R}$ is the average number of bonds per site.  The
couplings are i.i.d.r.v. with gaussian distribution with average
$J_0/c$ and variance $J/\sqrt{c}$,
\begin{equation}
  p(J_{ij})=\frac{1}{\sqrt{2\pi J^2/c}}
  \exp\left[-\frac{\left(J_{ij}-J_0/c\right)^{2}}{2J^2/c}\right].
  \label{eq21}
\end{equation}
The local random field $h_i$ are i.i.d.r.v. that follow a
a gaussian distribution
\begin{equation}
  p(h_i)=\frac{1}{\sqrt{2\pi\Delta^2}}
  \exp\left[-\frac{\left(h_i-h_0\right)^2}{2\Delta^2}\right].
  \label{eq22}
\end{equation}
with average $h_0$ and variance $\Delta$.

As usual, the thermal equilibrium properties are derived from the
free-energy
\begin{equation}
  f(\beta)=-\lim_{N\rightarrow\infty}\frac{1}{\beta N}\langle\log
  Z\rangle_{\{J_{ij},h_i,c_{ij}\}}\\,
  \label{freen}
\end{equation}
where the brackets stand for the disorder average and
$Z=\sum_{\mbox{\boldmath$\sigma$}}{\mathrm e}^{-\beta H}$ is the
partition function. The symbol $\mbox{\boldmath$\sigma$}$ represents a
$N$-coordinate system's state vector.

In order to average over the quenched disorder we follow the replica
method, where we need to calculate the average over the replicated
partition function instead of the logarithm of the partition
function. The replicated partition function becomes
\begin{align}
  \left\langle Z^n\right\rangle_{\{J_{ij},h_i,c_{ij}\}} =
  \sum_{\mbox{\boldmath$\sigma$}^1\cdots\mbox{\boldmath$\sigma$}^n}\Bigl\langle
  \exp\Bigl(\beta\sum_{i,\alpha}
  h_i\sigma_i^\alpha\Bigr)\exp\Bigl(\beta\sum_{i,j<i}J_{ij}c_{ij}
  \mbox{\boldmath$\sigma$}_i\cdot\mbox{\boldmath$\sigma$}_j\Bigr)
  \Bigr\rangle_{\{J_{ij},h_i,c_{ij}\}}\,.
  \label{rpart3}
\end{align}
The $N$-dimensional vector $\mbox{\boldmath$\sigma$}^\alpha$
represents the state of the whole network in the replica $\alpha$,
while the $n$-dimensional vector $\mbox{\boldmath$\sigma$}_i$
represents the state of the $n$ replicas in the site $i$.

The main outcome of the RS solution is a recursive equation for the
distribution of effective local fields (see the Appendix)
\begin{align}
  \label{spW}
  W(x)=\sum_k\frac{\mathrm{e}^{-c}c^k}{k!}  &
  \Bigl\langle\int\prod_{l=1}^k\,dx_l\,W(x_l) \\ &\times\delta\Bigl(x
  -h_l-\frac{1}{\beta}\sum_l\arctan\Bigl[\tanh\Bigl(\beta
    x_l\Bigr)\tanh\Bigl(\beta
    J_l\Bigl)\Bigr]\Bigr)\Bigr\rangle_{J_l,h_l}\,, \nonumber
\end{align}
that can be solved through a population dynamics algorithm to be
explained below. By knowing $W(x)$, the order parameters can be
obtained: the magnetization
\begin{align}
  m=\sum_{\sigma^1\cdots\sigma^n}P\bigl(\mbox{\boldmath$\sigma$}\bigr)
  \sigma^1 =\int dx\,W(x)\tanh\bigl(\beta x\bigr)\,.
\end{align}
and the spin-glass order parameter,
\begin{align}
  q=\sum_{\sigma^1\cdots\sigma^n}P\bigl(\mbox{\boldmath$\sigma$}\bigr)
  \sigma^1\sigma^2 =\int dx\, W(x)\tanh^2\bigl(\beta x\bigr)\,.
\end{align}

A key point concerns the stability of the RS solution. In fully
connected networks, the so called AT line is the locus where the
replicon eigenvalue vanishes \cite{AT}, but it becomes very difficult
to apply this method to finite connectivity networks. Here we follow
the method of two replicas \cite{TwoReplica,Neri}, that consists in calculating
the joint distribution
\begin{align}
  W(x,y)&=\sum_k\frac{\mathrm{e}^{-c}c^k}{k!}
  \Bigl\langle\int\prod_{l=1}^k\,dx_l\,dy_l\,W(x_l,y_l)
  \\ &\times\delta\Bigl(x -h_l
  -\frac{1}{\beta}\sum_l\arctan\Bigl[\tanh\Bigl(\beta
    x_l\Bigr)\tanh\Bigl(\beta J_l\Bigl)\Bigr]\Bigr) \nonumber
  \\ &\times\delta\Bigl(y -h_l
  -\frac{1}{\beta}\sum_l\arctan\Bigl[\tanh\Bigl(\beta
    y_l\Bigr)\tanh\Bigl(\beta J_l\Bigl)\Bigr]\Bigr)
  \Bigr\rangle_{J_l,h_l}\,.\nonumber
  \label{two}
\end{align}
When the RS solution is stable, the two replicas are identical, and
$W(x,y)$ is diagonal. When the RS is unstable, ergodicity is broken,
and the two-replica distribution is no longer diagonal. To localize
the edge of stability, it is easier to calculate the overlap between
two replicas,
\begin{align}
  q'=\int dx\,dy\, W(x,y)\tanh\bigl(\beta x\bigr)\tanh\bigl(\beta
  y\bigr)\,.
\end{align}
It occurs that $q'=q$ if RS is stable, and $q'\neq q$ otherwise.

\section*{III. Results}

\begin{figure}
  \begin{center}
    \includegraphics[width=8cm,clip]{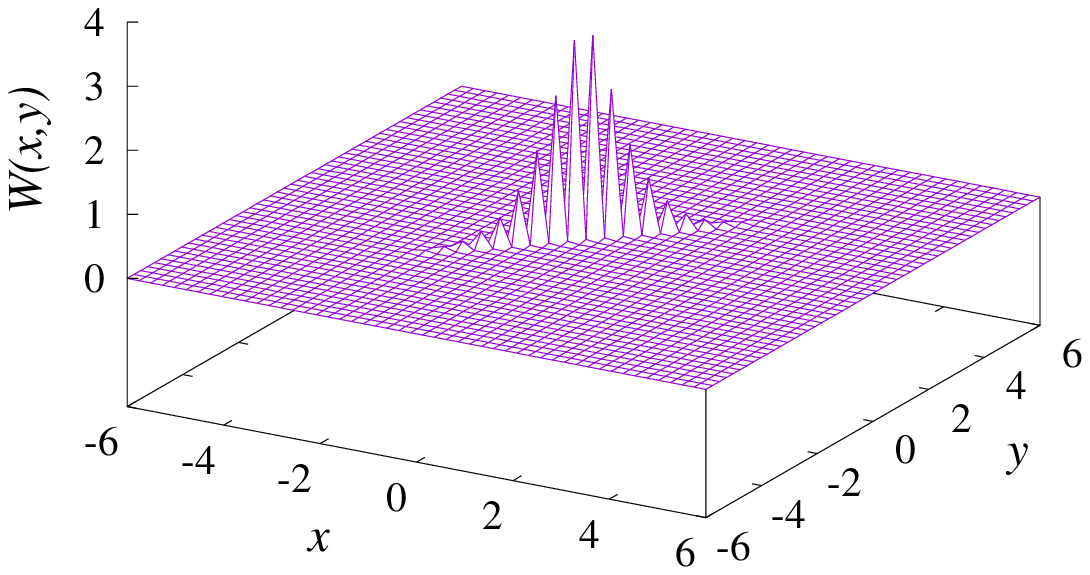}
  \end{center}
  \begin{center}
    \includegraphics[width=8cm,clip]{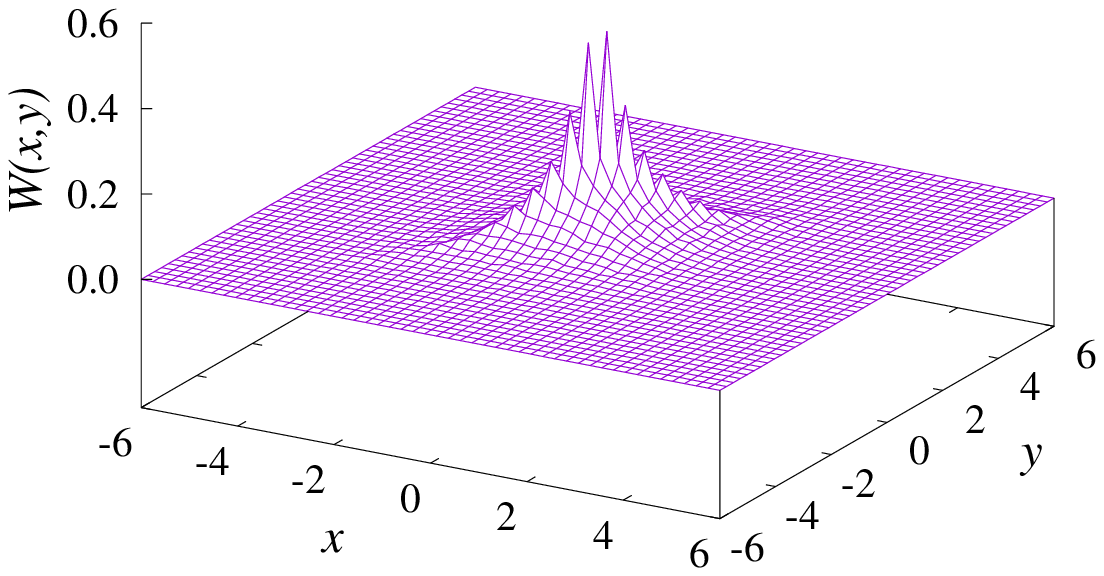}
  \end{center}
  \caption{Joint distributions for $c=4$, $h_0/J=0.2$ and
    $\Delta/J=0.0$. Top: $T/J=0.8$, PM phase, RS-stable diagonal distribution. Bottom: $T/J=0.6$, SG phase, RS-unstable non-diagonal distribution.}
  \label{figure1}
\end{figure}

The saddle-point equation for the local field distribution,
Eq. (\ref{spW}), is solved by the population dynamics method
\cite{skantzos}. It starts with a randomly chosen population of local
fields. Typically, the size of the population is 100,000 fields. The
method is iterative. Each iteration, a number $k\in\mathbb{N}$ is
sorted according to the poissonian distribution with average $c$. Then,
$k$ fields are randomly chosen from the field population and the
summation of the argument of the $\delta$-function in
Eq. (\ref{spW}) is evaluated. The result is assigned to another field
randomly chosen from the same population. This recipe is applied till
$W(x)$ converges. It takes, on average, 100 iterations per field to
converge. The distribution $W(x,y)$ is calculated similarly. Examples
of joint distributions are shown in Figure \ref{figure1}.

In the PM phase the system is ergodic, and the RS solution is
stable. As mentioned above, the correspondent joint distribution is
diagonal, as is shown in the top panel of Figure \ref{figure1}.
Conversely, the ergodicity is broken in the SG phase, i.e., this phase
is RS unstable, and the joint distribution is no longer diagonal, as
can be seen in the bottom panel of Figure \ref{figure1}. We proceed by
considering that the SG to PM transition coincides with the AT line.

\begin{figure}
\begin{center}
  \includegraphics[width=7cm,clip]{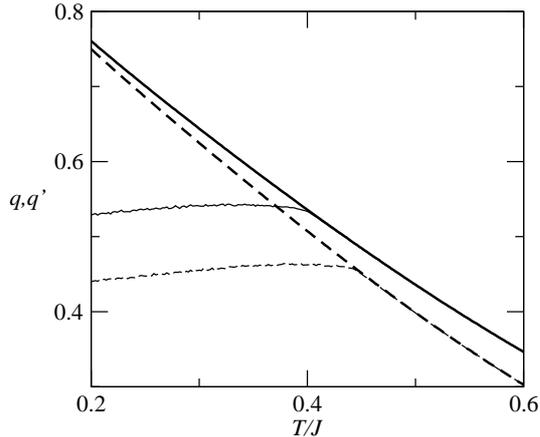}
\end{center}
  \caption{$q$ (thick lines) and $q'$ (thin lines) vs. $T/J$ for
    $c=4$, $h_0=0.15$, $\Delta=0$, $J_0=0.5$ (solid lines) and
    $h_0=0$, $\Delta=0.2$ and $J_0=0$ (dashed lines).  }
  \label{figure2}
\end{figure}

Representative examples of how the order parameters $q$ and $q'$
behaves as the temperature varies is shown in Figure \ref{figure2},
for $c=4$ and two sets of parameters $(h_0,\Delta,J_0)$. The set
$(0.15,0,0.5)$, representing an example of uniform magnetic field and
random couplings with a ferromagnetic constant, is shown in solid
lines.  The set $(0,0.2,0)$, representing an example fully disordered
magnetic local fields and couplings, is shown in dashed lines. This
figure reveals that the whole behavior is robust against changes on
the parameters, with ergodicity being broken at low temperature.

\begin{figure}
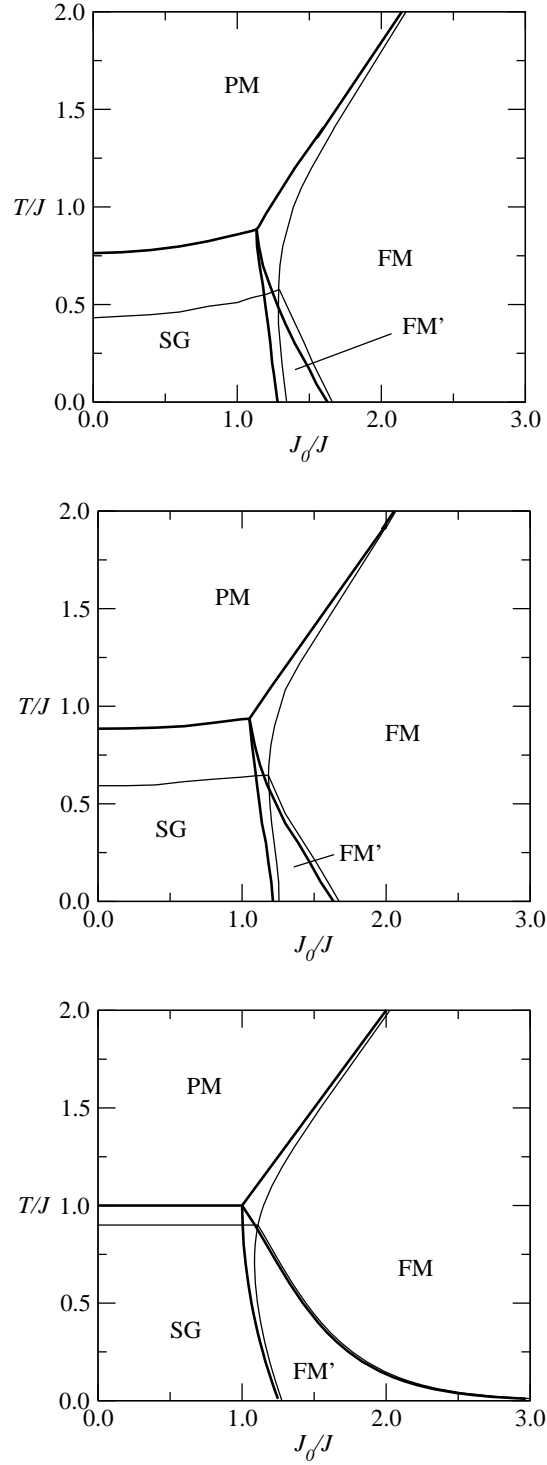

  \begin{center}
    \includegraphics[width=7cm,clip]{TJ0_4.eps} \hspace{0.5cm}
  \end{center}
  \begin{center}
    \includegraphics[width=7cm,clip]{TJ0_8.eps}
  \end{center}
  \begin{center}
    \includegraphics[width=7cm,clip]{TJ0_sk.eps}
  \end{center}
  \caption{$T/J$ vs. $J_0/J$ phase diagrams for $h_0/J=0.0$, $\Delta/J=0.0$ (thick lines) and $\Delta/J=0.2$ (thin lines), for $c=4$ (top), $c=8$ (middle) and fully connected (bottom).}
  \label{figure3}
\end{figure}

A more complete view of the role played by the random field and average connectivity on finite connectivity spin-glasses is revealed through the phase diagrams. In Figure \ref{figure3}, $T/J$ vs. $J_0/J$ phase diagrams for $c=4$, $c=8$ and the fully connected network, with random field and without random field are presented.  
The SG to PM transition, as well as the transition from the mixed phase FM' to FM are AT lines, i.e., lines that signal the locus where the ergodicity is broken, with $q$ and $q'$ becoming different. 
The FM' to SG and F to PM transitions are signaled by the magnetization $m$ going to zero. All the transitions are continuous. The mixed FM' phase is a non-ergodig ferromagnetic phase, where $m\neq 0$, $q>0$ (it is the correlation of a replica with itself), $q'\neq 0$ (it is the correlation between two distinct replicas), and $q\neq q'$.
As a general remark, when increasing the random field the transition lines are displaced in a way that the surface occupied by more entropic phases increases, and the less entropic phases are reduced. 
The differences between averages connectivies $c=4$ and $c=8$ are mainly quantitative. When increasing $c$ the transition lines are again
displaced, this time in a way the area occupied by the less entropic phases increases, while the more entropic are reduced. So speaking,
the SG to PM transition line displaces upwards, the SG to FM' displaces to the left and FM to PM displaces to the left and FM to FM' displaces downwards.

It should be remarked that, by comparing finite and fully connectivity, the relevant qualitative difference is that, in the
fully connected case, the transition line FM' to FM goes asymptotically to $T=0$ when $J_0/J$ increases, unlike the finite case $c$, where the transition line intersects the $T$ zero axis. This means that a finite connectivity favors the ergodicity at zero temperature.

\begin{figure}
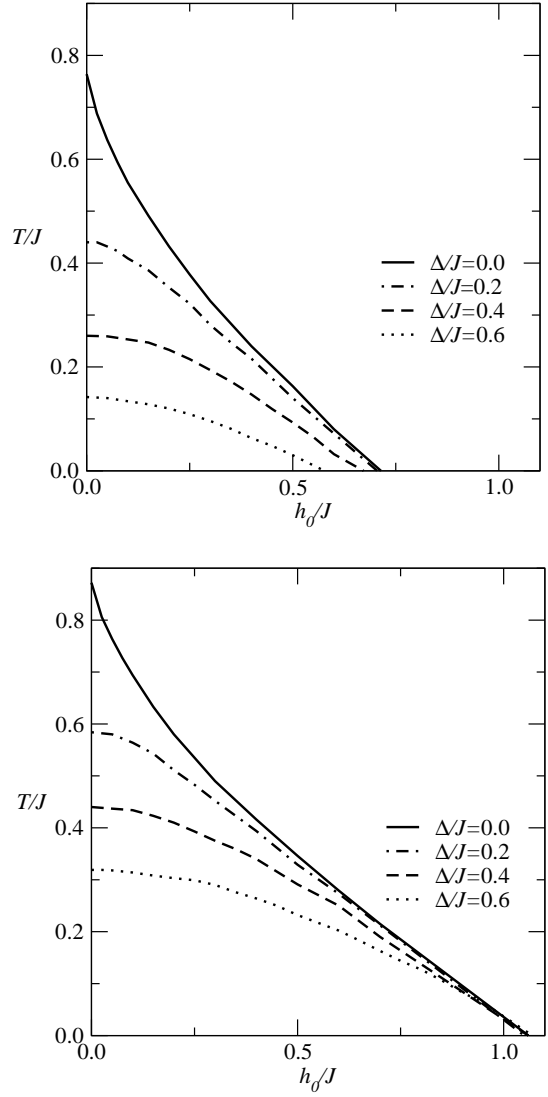

  \begin{center}
    \includegraphics[width=7cm,clip]{ThD_4.1.eps} \hspace{0.5cm}
  \end{center}
  \begin{center}
    \includegraphics[width=7cm,clip]{ThD_8.1.eps}
  \end{center}
  \caption{Top: $T/J$ vs. $h_0/J$ phase diagrams for $c=4$ and $J_0=0$
    for different values of $\Delta/J$. Bottom: the same, but
    for $c=8$.}
  \label{figure4}
\end{figure}

Next, we analyze the SG to PM transition through $T/J$ vs. $h_0/J$
phase diagrams. Figure \ref{figure4} shows the SG to PM transitions for
connectivity $c=4$ and $c=8$, for four representative values of the
field disorder. Although the pictures for different values of $c$ are
qualitatively similar, with both the uniform field component $h_0$ and
the random field component $\Delta$, both given in units of $J$,
acting to suppress the SG phase in favor of the PM phase, there are
some aspects to consider. First, $\Delta$ is much more effective in
suppressing the SG at small than at a large $h_0$. Second, it is also
more effective the smaller the $c$.  This, even considering that both
the mean value and the variance of the couplings scale with $c$ (see
Eq. \ref{eq21}).  This means that a more connected network with weaker
couplings produces a more robust SG phase than a less connected
network with stronger couplings one.  Other aspect that must be very
stressed is that the SG phase is suppressed completely above a certain
$h_0$ for any value of $\Delta$. Moreover, for the same fixed value of
$\Delta$, this suppression is much more effective for c=4 than for
c=8. 

\begin{figure}
  \begin{center}
    \includegraphics[width=7cm,clip]{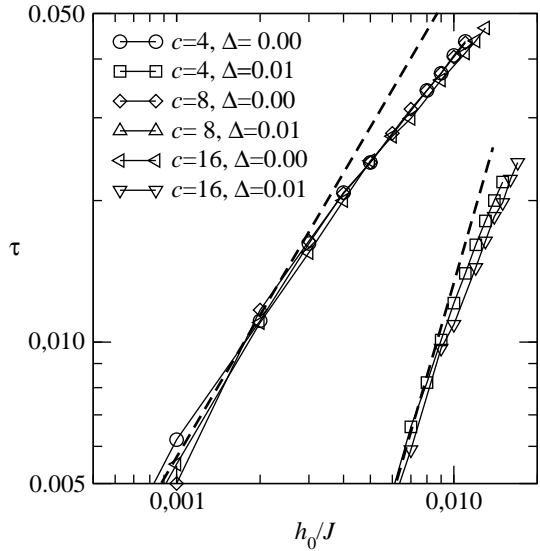}
  \end{center}
  \caption{Reduced temperature vs. RF amplitude
    $h_0/J$ for $\Delta/J=0.00$ and $\Delta/J=0.01$, $c=4$, $c=8$ and $c=16$. The solid lines are only guidelines. Shown in dashed, slope 1 and slope 2 straight lines.}
  \label{figure42}
\end{figure}

We remark in Figure \ref{figure4} that the convexity of the curves at
small $h_0/J$ values changes from $\Delta/J=0$ to
$\Delta/J>0$. To investigate this in detail, in Figure
\ref{figure42} we plot, in logarithmic scale, the reduced temperature $\tau=(T_0-T)/J$ vs. $h_0/J$, for small $h_0/J$, where $T_0=T(h_0/J=0)$, for $\Delta/J=0.00$ and $\Delta=0.01$, with connectivity ranging from $c=4$ to $c=16$. If the reduced temperature is expressed as a power law $\tau\sim(h_0/J)^{2/\phi}$ as $h_0/J\rightarrow 0$, the slopes in figure indicate that 
$\phi=2$ for $\Delta/J=0.00$ and $\phi=1$ for $\Delta/J$ as small as 0.01.
Here, we compare our results with those obtained for the fully connected network \cite{Soares1994}, where 
$\phi=3$ for  $\Delta/J=0.00$, identified as SG regime, and 
$\phi=1$ for finite $\Delta/J$, identified as the RF regime. We guess that the $\phi=3$ in the SG regime is a particularity of $c\rightarrow\infty$, since the curves for increasing $c$ in Figure \ref{figure42} superimpose, 
suggesting that $\phi=2$ in the SG regime is robust for all finite $c$.
To resume, the results for both finite and fully connected networks
indicates that a crossover from SG to RF regime takes place as $\Delta/J$ becomes non
zero. 

\begin{figure}
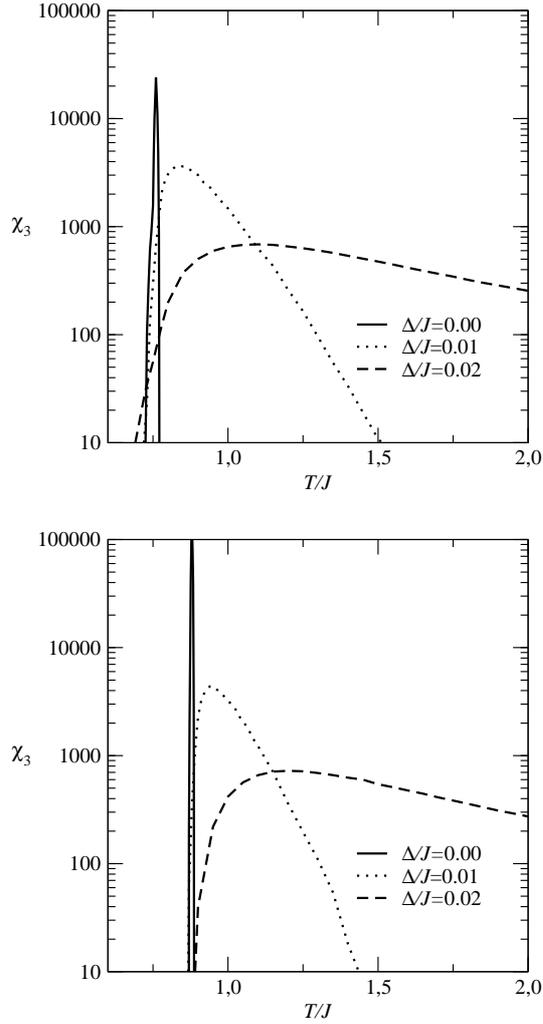

    \centering
    \includegraphics[width=7cm,clip]{sus4log.eps}
    
    \vspace{0.5cm}
    \includegraphics[width=7cm,clip]{sus8log.eps}
    \caption{Top: nonlinear susceptibility vs. temperature for $c=4$ and several values of $\Delta/J$. Bottom: the same, but for $c=8$.}
    \label{sus}
\end{figure}

\begin{figure}
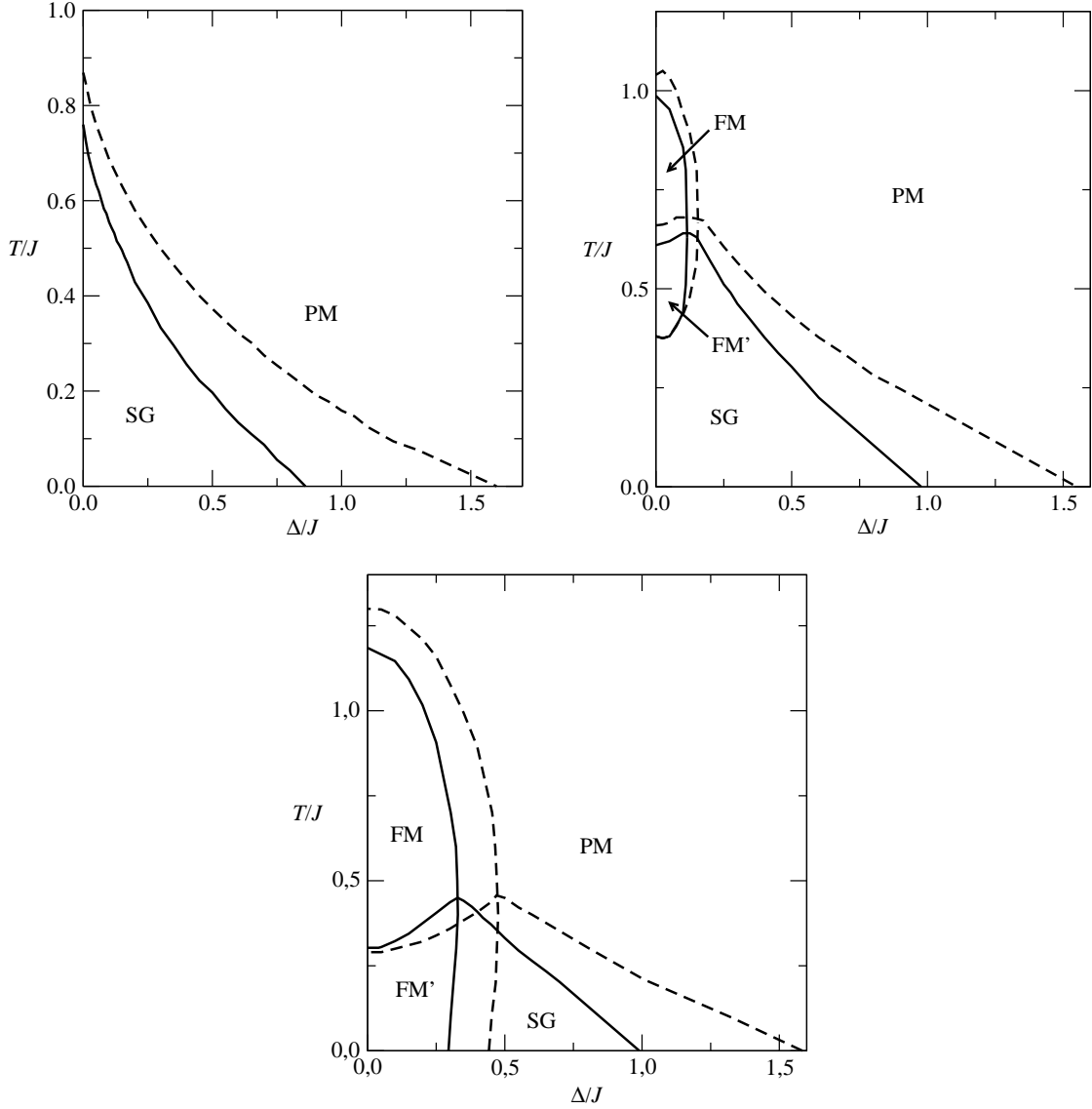

  \begin{center}
    \includegraphics[width=7cm,clip]{Tvth0.eps}\quad\quad
    \includegraphics[width=6.9cm,clip]{T_vth.eps}
  \end{center}
  \begin{center}
    \includegraphics[width=7cm,clip]{Tvth01.4.eps}
  \end{center}
  \caption{$T/J$ vs. $\Delta/J$ phase diagram for $h_0=0$. Top left: $J_0/J=0$; $c=4$ (solid lines) and $c=8$ (dashed lines). Top right: $J_0/J=1.22$ and $c=4$ (solid lines); $J_0/J=1.15$ and $c=8$ (dashed lines). Bottom: $J_0/J=1.4$; $c=4$ (solid lines) and $c=8$ (dashed lines). All the transitions are continuous.}
  \label{figure5}
\end{figure}

\begin{figure}
  \begin{center}
    \includegraphics[width=7cm,clip]{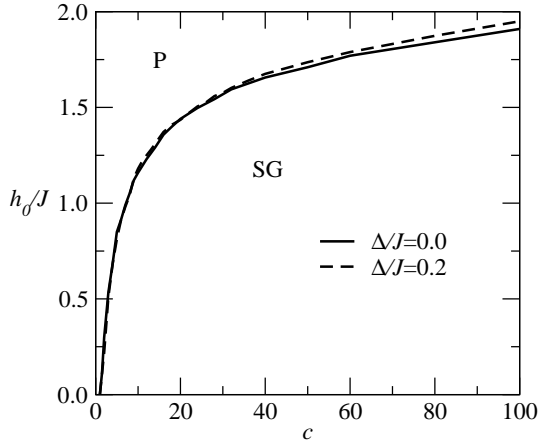}
  \end{center}
  \caption{Limiting $h_0/J$ of the SG phase vs. $c$ at $T=0$, for
    $\Delta/J=0.0$ and $\Delta/J=0.2$. Below and above the lines are
    the loci of the SG (non-ergodic) and PM (ergodic), respectively.}
  \label{figure6}
\end{figure}

The non-linear susceptibility $\chi_3=\partial^3 m/\partial h_0^3|_{T/J,h_0/J=0}$ diverges at the SG to PM transition (AT line) in the SG regime (zero $\Delta/J$) \cite{binderyoung}. The non-linear susceptibility  as a function of $T/J$ is shown in Figure \ref{sus}, for some small values of $\Delta/J$. For $\Delta/J=0$ indeed there is a divergence at the same $T/J$ where the two-replica method localizes the SG to PM transition, for both $c=4$ and $c=8$. For $\Delta/J>0$ there is a peak instead of a divergence, that no longer signals the SG to PM transition. As $\Delta/J$ increases, the peak becomes quickly less pronounced and moves to higher temperature values, as can be seen in Figure \ref{sus}. 

To complete the description, $T/J$ vs. $\Delta/J$ phase diagrams for $h_0/J=0$ and $J_0/J$ constant are present in Figure \ref{figure5}. In the top left panel we have $J_0/J=0$, that is a prototype for all phase diagrams where the uniform part of the coupling constant is to weak to allow the appearing of ferromagnetic phases. In the top right panel, a constant $J_0/J$ in the re-entrant region was chosen. Here, the uniform coupling becomes sufficiently strong to allow the appearing of FM and FM' phases, although the SG remains as the most ordered phase at zero temperature. For $c=4$ and $c=8$ this takes place, e.g., in the neighborhood
of $J_0/J=1.22$ and $J_0/J=1.15$, respectively. In the bottom panel the $T/J$ vs. $\Delta/J$ phase diagram for a constant $J_0/J=1.4$ for both $c=4$ and $c=8$ is shown. Here, the uniform coupling becomes stronger, and the mixed phase FM' can be found at zero temperature.
The phase diagrams are qualitatively
similar for both values of the connectivity. All the transitions are
continuous. As $\Delta/J$ increases, FM and FM' phases are the first to
be suppressed. Then, the SG to PM transition line decreases
monotonically to $T/J=0$ and PM is the only remaining phase at large
$\Delta$.

The last comment concerns the comparison with the fully connected
network. Here again, the FM' to FM transition intercepts the zero
temperature axis, contrary to the fully connected network. If the
two-replica method to localize the AT line is correct, the finite
connectivity results should approach the fully connected ones as $c$
increases. This seems to be the case for the general aspects of the
phase diagrams shown above, except in the high field and high coupling
constant regimes. In the fully connected network there is no
paramagnetic phase at zero temperature, in contrast to the finite
connectivity case. Note that this does not mean that there is no
magnetization at zero $T$: there is, indeed a field-induced
magnetization. The main outcome is that we can find the
finite connectivity network ergodic at zero $T$, contrary to the fully
connectivity network. To illustrate how the zero $T$ ergodicity region
evolves as a function of the connectivity is shown in Figure
\ref{figure6}. The figure shows the limiting $h_0/J$ of the SG phase, at
$T=0$. As one should expect, this limit increases slowly but
monotonically with $c$. 

\section{Conclusions}

In this work we have investigated the problem of the joint
presence of SG and RF using 
random network.
In order to
accomplish that, we have considered the Random Network Random Field
Ising Model \cite{Doria} where the spin exchange
interaction as well as the RF are random variables following a
gaussian distribution.  Our goal has been, using the connectivity as a
control parameter in the theory, to verify whether and which are the
differences 
with the mean field theory 
i.e., the fully connected theory
\cite{Soares1994}.  Particularly, in the
presence of a magnetic field. As a methodological
novelty in the problem, we performed the check of the
stability of the RS 
solution using the Two-replica
method.  This procedure, which gives the AT line,  
has been used for the SG problem
without RF.  Thus, in fact, it can be considered that our work 
also belongs to a more general discussion
concerning the description of the SG non-trivial ergodicity breaking
using the AT line when the network connectivity can vary.

Following a population dynamics algorithm, the effective distribution
of local fields was determined, allowing to the calculation of
relevant order parameters. Then, we obtained phase diagrams
temperature versus ferromagnetic exchange interaction $J_0$ (see
Eq. (\ref{eq21})) and temperature versus the magnetic field $h_0$ (see
Eq. (\ref{eq22})) for several values of the random field variance
$\Delta$ and for two values of the average connectivity, namely $c=4$
and $c=8$.  All energy scales in the problem are given in units of the
variance $J$ of the random exchange spin interaction. The differences
in the phase diagrams with the two values of $c$ are mainly
quantitative.  Nevertheless, the results have shown that the less
entropic phases occupy growing areas to the detriment of the more
entropic ones as $c$ increases. This means that the connectivity
favors the ordered phases, even considering that the coupling constant
is correctly normalized with $c$ (see Eq. \ref{eq21}). Moreover, we do
remark that the AT line intercepts both the $J_0$ and $h_0$ axis at
zero temperature, contrary to the observed in the fully connected theory. 
In other words, the SG ground state prevails only within a
certain interval of $J_0$ and $h_0$. This also means that, even
with quenched disorder, for finite connectivity there is a region at
zero temperature where the ergodicity remains unbroken above a given
value of the magnetic field, no matter the strength of the RF gaussian
variance $\Delta$. We notice that, in the limit of large values of
$c$, there are indications that the 
fully connected theory
is recovered,
particularly with regard to the 
AT line.

To conclude, one of the main outcomes of the present investigation concerns the
crossover between the RF and the SG regime.  Within the fully
connected theory, the crossover between the RF and SG regimes was
described by $\tau \equiv T_0-T\sim h_0^{2/\phi}$ ($T_0$ is the
freezing temperature without field). In the fully connected theory, the values $\phi=1$ and $3$
corresponds to RF and SG regimes, respectively.  
We found, in this work, that at small $h_0$, $\tau\sim h_0$ at $\Delta=0$ and $\tau\sim h_0^2$ for any finite $\Delta$. In other words, $\phi=1$ and $\phi=2$ in the RF and SG regimes, respectively.

\section*{Acknowledgments}

The authors acknowledge F. L. Metz and F. D.  Nobre for fruitfull
discussions. The present work was supported, in part, by the Brazilian
agency CNPq.

\section*{Appendix}

The finite connectivity replica method has become standard. We rewrite
here only some key points and refer to \cite{Mo98,ETM} for
details. After averaging over $c_{ij}$ we obtain, in the $c/N\rightarrow
0$ limit,
\begin{align}
  \left\langle Z^n\right\rangle_{\{J_{ij},h_i,c_{ij}\}}
  =\sum_{\mbox{\boldmath$\sigma$}^1
    \cdots\mbox{\boldmath$\sigma$}^n}\Bigl\langle
  \exp\Bigl[\beta\sum_{i,\alpha} h_i\sigma_i^\alpha
    +\frac{c}{2N}\sum_{i,j\neq i}\Bigl(\mathrm{e}^{\beta J_{ij}
      \mbox{\boldmath$\sigma$}_i\cdot\mbox{\boldmath$\sigma$}_j}-1\Bigr)\Bigr]
  \Bigr\rangle_{\{J_{ij},h_i\}}\,.
  \label{rpart4}
\end{align}
Next, we introduce the fraction $P(\mbox{\boldmath$\sigma$})$ of sites
where the replica configuration $\mbox{\boldmath$\sigma$}$ is realized
and the auxiliary variables $\hat{P}(\mbox{\boldmath$\sigma$})$ and
evaluate the trace over the spin variables. This reduces to the
problem of one site, and the replicated partition function can be
rewritten as
\begin{align}
\nonumber \left\langle Z^n\right\rangle_{\{J_{ij},h_i,c_{ij}\}} &
=\int\prod_{\mbox{\boldmath$\sigma$}}dP\left(\mbox{\boldmath$\sigma$}\right)
d\hat{P}\left(\mbox{\boldmath$\sigma$}\right)
\exp\Bigl\{N\log\sum_{\mbox{\boldmath$\sigma$}}\Big\langle\exp
\Big[\beta h\sum_\alpha\sigma^\alpha
  -\hat{P}\left(\mbox{\boldmath$\sigma$}\right)\Big]\Big\rangle_h\\ &+
N\sum_{\mbox{\boldmath$\sigma$}}\hat{P}\left(\mbox{\boldmath$\sigma$}\right)
P\left(\mbox{\boldmath$\sigma$}\right)
+\frac{Nc}{2}\sum_{\mbox{\boldmath$\sigma$}\mbox{\boldmath$\sigma$}'}
P\left(\mbox{\boldmath$\sigma$}\right)P\left(\mbox{\boldmath$\sigma$}'\right)
\Bigl\langle\Bigl(\mathrm{e}^{\beta
  J\mbox{\boldmath$\sigma$}\cdot\mbox{\boldmath$\sigma$}'}
-1\Bigr)\Bigr\rangle_J\Bigr\}\,.
  \label{rpart5}
\end{align}
In the limit $N\rightarrow\infty$, the saddle-point method
applies, and the free-energy becomes
\begin{align}
  f(\beta)=-\lim_{n\rightarrow 0}&\frac{1}{\beta
    n}\mathrm{Extr}_{P(\mbox{\boldmath$\sigma$})}\Bigl\{
  -\frac{c}{2}\sum_{\mbox{\boldmath$\sigma$}\mbox{\boldmath$\sigma$}'}
  P(\mbox{\boldmath$\sigma$})P(\mbox{\boldmath$\sigma$}')
  \Bigl\langle\Bigl(\mathrm{e}^{\beta
    J\mbox{\boldmath$\sigma$}\cdot\mbox{\boldmath$\sigma$}'}
  -1\Bigr)\Bigr\rangle_J\nonumber
  \\ &+\log\sum_{\mbox{\boldmath$\sigma$}}\Bigl\langle\exp\Bigl[ \beta
    h\sum_\alpha\sigma^{\alpha}+c\sum_{\mbox{\boldmath$\sigma$}'}
    P(\mbox{\boldmath$\sigma$}')\Bigl\langle\Bigl(\mathrm{e}^{\beta
      J\mbox{\boldmath$\sigma$}\cdot\mbox{\boldmath$\sigma$}'}
    -1\Bigr)\Bigr\rangle_J\Bigr]\Bigr\rangle_h\Bigr\}\,,
\label{el1}
\end{align}
where the auxiliary variables $\hat{P}(\mbox{\boldmath$\sigma$})$ were
eliminated by the saddle-point equations $\partial f(\beta)/\partial
P(\mbox{\boldmath$\sigma$})=0$.  The variables
$P(\mbox{\boldmath$\sigma$})$ must satisfy the remaining saddle-point
equations,
\begin{align}
P(\mbox{\boldmath$\sigma$})=\dfrac{\Bigl\langle\exp\Bigl[ \beta
    h\sum_\alpha\sigma^{\alpha}+c\sum_{\mbox{\boldmath$\sigma$}'}
    P(\mbox{\boldmath$\sigma$}')\Bigl\langle\Bigl(\mathrm{e}^{\beta
      J\mbox{\boldmath$\sigma$}\cdot\mbox{\boldmath$\sigma$}'}
    -1\Bigr)\Bigr\rangle_J\Bigr]\Bigr\rangle_h}{\sum_{\mbox{\boldmath$\sigma$}'}
  \Bigl\langle\exp\Bigl[ \beta
    h\sum_\alpha\sigma'^\alpha+c\sum_{\mbox{\boldmath$\sigma$}''}
    P(\mbox{\boldmath$\sigma$}'')\Bigl\langle\Bigl(\mathrm{e}^{\beta
      J\mbox{\boldmath$\sigma$}'\cdot\mbox{\boldmath$\sigma$}''}
    -1\Bigr)\Bigr\rangle_J\Bigr]\Bigr\rangle_h}\,.
\label{sp}
\end{align}

We are interested in those solutions satisfying the RS ``ansatz'',
\begin{align}
  P(\mbox{\boldmath$\sigma$})=\int dx\,W(x)\frac{\mathrm{e}^{\beta
      x\sum_{\alpha}\sigma_{\alpha}}}{[2\cosh(\beta x)]^n}\,.
  \label{ansatz}
\end{align}
This expression is equivalent under permutation of replicas.
Introducing the RS ansatz in Eq. (\ref{sp}), we obtain a recursive
equation for the distribution of effective local fields $W(x)$,
Eq. (\ref{spW}).

\end{document}